\documentclass[a4paper,11pt]{article}
\pdfoutput=1 

\usepackage{jheppub}

\usepackage[mmddyyyy,hhmmss]{datetime}

\usepackage{amsmath,graphicx,amssymb,subfigure,bbm,comment}
\usepackage[all]{xy}

\allowdisplaybreaks

\title{Giant magnons and spiky strings in the Schr\"odinger/ dipole-deformed CFT correspondence}

 \author{George Georgiou$^a$,}
\author{Dimitrios Zoakos$^b$}

\affiliation{$^a$ Institute of Nuclear and Particle Physics, National Center for Scientific Research Demokritos, 15310 Athens, Greece}
\affiliation{$^b$ Centro de F\'isica do Porto, Universidade do Porto, Rua do Campo Alegre 687, 4169-007 Porto, Portugal.}

\emailAdd{georgiou@inp.demokritos.gr}
\emailAdd{zoakos@gmail.com}

\abstract{We construct semi-classical string solutions of the Schr\"odinger $Sch_5 \times S^5$ spacetime, which is
conjectured to be the gravity dual of a non-local dipole-deformed CFT. They are the counterparts of the giant magnon and 
spiky string solutions of the undeformed $AdS_5 \times S^5$ to which they flow when the deformation parameter is turned off. 
They live in an $S^3$ subspace of the five-sphere along 
the directions of which the $B$-field has non-zero components having also extent in the $Sch_5$ part of the metric. Finally, we speculate on the form of the dual field theory operators.}

\begin{document}
\maketitle
\flushbottom


\section{Introduction}

Integrability is the powerful tool behind the recent advances in understanding the gauge-gravity correspondence in the 
planar limit (for a review see \cite{Beisert:2010jr}). On the gauge theory side, diagonalizing the matrix of the 
anomalous dimensions can be performed through Bethe ansatz techniques \cite{Minahan:2002ve} since the anomalous dimension matrix can be mapped to the Hamiltonian of an integrable spin chain.
The magnons play the role of the asymptotic states and are considered as the fundamental single particle excitations  propagating on a certain BMN vaccum. Integrability of the theory implies that any 
scattering process factorizes in a scattering between two magnons.  As a result, the role of integrability is instrumental in determining the spectrum of ${\cal N}=4$ SYM for any value of the coupling \footnote{Recently, progress has also been made in the calculation of higher point correlation functions using integrability. For an incomplete list of work in this direction see \cite{3-point} and references therein.}.

On the string side, the strong coupling dual to the field theory magnons are the so-called giant magnons \cite{Hofman:2006xt, Minahan:2006bd, Chen:2006gea, Benvenuti:2008bd, Spradlin:2006wk}. These are semi-classical solutions  of string theory on $AdS_5\times S^5 $ and are described by open strings moving in a subspace 
of the sphere $S^5$ with finite angular extent. Their dispersion relation is the strong coupling limit  
of the exact dispersion relation obtained from the gauge theory analysis \cite{Beisert:2005tm}.  
Similar classical string configurations with finite angular amplitude that wind infinitely many times around the 
orthogonal angular direction are called single spiky strings, \cite{Ishizeki:2007we, Ishizeki:2007kh}. 
Their gauge theory interpretation needs further clarification.

Over the last years there is a lot of attention on the integrable deformations of the AdS/CFT. Two important examples of such deformations are the 
$\beta$ deformation \cite{Leigh:1995ep,Lunin:2005jy} for which the magnon solution was calculated in \cite{Chu:2006ae} and  the non-commutative theories \cite{Seiberg:1999vs}. Recent activity is coming
from the null dipole deformation of the ${\cal N}=4$ SYM, which is special since it has minimal non-locality along one of its light-like directions.
The dual gravity background is known as a  Schr\"odinger space-time, it was first worked out in \cite{Alishahiha:2003ru} and can be obtained by performing an abelian T-duality along one of the isometries of the five-sphere followed 
by a shift along a light-like direction of the $AdS_5$ boundary and a second T-duality along the dual coordinate of the sphere. On the field theory side, the deformation is realised by introducing the appropriate $\star$-product among the fields of the $\mathcal{N}=4$ Lagrangian.
Generically, the $\star$-product used definition of the deformed theory can be  identified with a corresponding Drinfeld-Reshetikhin twist of the underlying integrable structure \cite{Beisert:2005if,Ahn:2010ws}.

Integrability issues of the dipole-deformed CFT were discussed in detail in \cite{Guica:2017mtd}\footnote{Integrability 
of the $\sigma$-model at the classical level was discussed in 
\cite{Kawaguchi:2014qwa,vanTongeren:2015uha,Matsumoto:2015uja,Kameyama:2015ufa}, where the Schr\"odinger spacetime was
realized as a particular Yang--Baxter deformation of the $AdS_5 \times S_5$ superstring.}. The deformation is realized
as a Drinfeld-Reshetikhin twist \cite{Drinfeld:1989st,Reshetikhin:1990ep} of the ${\cal N}=4$ integrable structure. 
In  \cite{Guica:2017mtd} the authors provided a test of the Schr\"odinger holography by matching the
anomalous dimensions of long gauge theory operators with the strong coupling prediction of certain BMN-like strings (see also \cite{Ouyang:2017yko}). 
While the present paper was being written the work  \cite{Ahn:2017bio} appeared. The authors have calculated the dispersion relation of a giant magnon-like solution in $Sch_5\times S^5$. However, their solution is fundamentally different from ours since it is point-like in the 
Schr\"odinger space-time part of the metric unlike the ones presented here \footnote{In addition to this the dispersion relation of \cite{Ahn:2017bio} 
when expressed solely in terms of the conserved quantities takes the from 
$\frac{(E-\sqrt{\mu^2 S^2+J_+^2+4 J_{\phi}^2})^2 E^2}{E^2-(\mu^2 S^2+J_+^2)}=T^2 \sin^2\frac{p}{2}$ 
which is completely different from ours \eqref{dispersionGM-final-1}}. 

In this paper we construct semi-classical string solutions living in the Schr\"odinger $Sch_5 \times S^5$ spacetime, 
which is conjectured to be the gravity dual of a certain non-local dipole-deformed CFT. 
Those solutions are the counterparts of the giant magnon and the single spike of the undeformed $AdS_5 \times S^5$, 
since in the zero deformation limit they become the ordinary giant magnon and single spike of the original 
$AdS_5 \times S^5$ background. 
We find those solutions in an $S^3$ subspace of the five-sphere along the direction of which the $B$-field has 
non-zero components. Furthermore, the solutions presented here are not point like but 
also extend in the $Sch_5$ part of the metric. For the dyonic giant magnon the dispersion 
relation is given by 
\begin{equation}
\label{dispersionGM-final-1}
\sqrt{E^2 \, - \, \frac{\lambda}{4 \, \pi^2} \, L^2 \, M^2} \, - \, J_1\,=\,
\sqrt{J_2^2 \, + \, \frac{\lambda}{\pi^2} \, \sin^2\frac{p}{2}}
\end{equation}
where the angle difference $\Delta \varphi_1$ is identified with the momentum  $p$ carried by the excitation, 
while for the single spike is given by
\begin{equation}
\label{dispersionSS-final}
 \, J^2_1 \, - \, J^2_2
\,=\, \frac{\lambda}{\pi^2} \, 
\sin^2\left[\frac{ \pi}{\sqrt{\lambda}} \, E \, - \, \frac{1}{2}\, L \, M  \, - \, \frac{1}{2} \, \Delta \varphi_1\right] \, .
\end{equation}

The paper is organised as follows: In section \ref{GM&SS} we present the ansatz for the solution and discuss the 
boundary conditions corresponding either to the giant magnon or to the single spike case.  After writing the solution 
we calculate the angular amplitudes. In section \ref{GM} we focus on the giant magnon case by finding the finite combinations
of the conserved charges and extracting the dispersion relation. In section \ref{SS} we provide the finite combination of 
global charges and the dispersion relation for the single spiky string solution. In section \ref{discussion}
we summarise and discuss about the possible form of the field theory operators that are dual to the giant magnon solutions 
in this dipole-deformed CFT. Furthermore, we comment on a possible, exact in $\lambda$, dispersion relation of the giant magnon solution. Our proposed relation \eqref{dispersionGM-final-2} when expanded agrees, in the BMN limit, with the one-loop result that is obtained by solving the Baxter equation.  Finally, an appendix contains the details of the 10-dimensional background as well as the proof that our solution lives in a consistent truncation of the full background.


\section{Giant magnons and single spiky strings on $Sch_5\times S^3$}
\label{GM&SS}

We consider the following truncation of the 10d $Sch_5\times S^5$ metric\footnote{We 
have set $\ell=1$ and $\alpha'=1$.} on the sphere $S^3$, 
which as we explain in appendix \ref{Polyakov-truncation} is consistent
\begin{equation} \label{globalsch5s3}
ds^2 = - \left(1+ \frac{\mu^2}{Z^4} \right) dT^2  +  \frac{1}{Z^2} \, \left(2 dT dV + dZ^2 \right) + 
\frac{1}{4}\,\Big[d\theta^2+\sin^2\theta d\phi^2+\left(d\psi - \cos\theta d\phi \right)^2 \Big]
\end{equation}
where $\psi\in[0,4\pi)$, $\theta\in[0,\pi]$ and $\phi\in [0,2\pi)$ are the angular coordinates of the three 
dimensional sphere and their ranges.
The B-field is given by the following expression 
\begin{equation} \label{Bfield}
B \, = \, \frac{\mu}{2\, Z^2} \, d T \wedge \left(d\psi - \cos\theta d\phi \right). 
\end{equation}
In order to compare the dispersion relation of the magnon with the standard result from the literature \cite{Hofman:2006xt}, 
we have to perform the following change of variables
\begin{equation} \label{S3coord}
\theta\,=\,2\,\eta
\hspace{1.5cm}
\psi\,=\,\varphi_1+\,\varphi_2
\hspace{1.5cm}
\phi\,=\,\varphi_1-\,\varphi_2
\end{equation} 
where $\eta \in [0,\pi/2]$, $\varphi_1\in [0,2\pi)$ and $\varphi_2\in [0,2\pi)$ are the ranges of those variables.

We now introduce the following ansatz to find classical solutions
\begin{eqnarray} \label{ansatz}
&& 
T\,=\, \kappa \,\tau \, + \, T_y(y)
\hspace{1.5cm}
V \,=\,\alpha \, \tau \,  + \, V_y(y) 
\hspace{1.5cm}
Z \,=\, Z_0
\nonumber \\ [3pt]
&& \theta\,=\,\theta_y(y)
\hspace{1.5cm}
\psi\,=\,\omega_\psi\,\tau\,+\,\Psi_y(y)
\hspace{1.5cm}
\phi\,=\,\omega_\phi\,\tau\,+\,\Phi_y(y)
\end{eqnarray} 
where we have defined the variable $y$ as  
\begin{equation} \label{definition-y}
y\, \equiv \, c\,\sigma-d\,\tau \, . 
\end{equation}
Additionally $\kappa$, $\alpha$, $Z_0$, $\omega_\psi$ and $\omega_\phi$ are constants, 
while $T_y(y)$, $V_y(y)$, $\theta_y(y)$, $\Psi_y(y)$ and $\Phi_y(y)$ are functions of $\sigma$ and $\tau$ (in the combination 
$c\,\sigma-d\,\tau$) that we will determine through the equations of motion and the Virasoro constraints. 


\subsection{Equations of motion} 
\label{EOM}

Applying the ansatz \eqref{ansatz} on the equation of motion that is coming from the variation of \eqref{Polyakov} along the direction $V$, 
we end up with an equation for the function $T_y(y)$. It is possible to integrate this differential equation and fully determine the 
function
\begin{equation} \label{solutionT}
T_y(y) \, = \, \frac{y}{c^2-d^2} \, \left( A_V \, Z_0^2 - d \, \kappa \right)
\end{equation}
where $A_V$ is the integration constant.

The equation of motion coming from the variation of \eqref{Polyakov} along the direction $T$, written in terms of $y$, can be integrated once, providing $V_y'(y)$ in terms of $\theta_y(y)$ and constants as follows
\begin{equation} \label{constraintV}
V_y'(y) \, = \, \frac{1}{c^2-d^2} \, \Bigg[A_V \left(\mu ^2+ Z_0^4\right) - A_T \, Z_0^2 - \alpha \, d +
\frac{1}{2} \, c \, \mu  \, \left( \omega_{\psi} - \omega_{\phi}  \cos \theta_y \right) \Bigg]
\end{equation}
where $A_T$ is the integration constant.

In the same fashion, the equations of motion coming from the variation of \eqref{Polyakov} along the directions $\psi$ and $\phi$ 
(in terms of $y$) can be integrated to express $\Psi_y'(y)$ and $\Phi_y'(y)$ in terms of $\theta_y(y)$ and 
constants as follows
\begin{eqnarray} \label{constraintPsi}
\Psi_y'(y)& = &
\frac{1}{c^2-d^2}
\left[\,\frac{4}{\sin^2\theta_y}\left(A_\psi+A_\phi \cos\theta_y \right)
 - d\,\omega_\psi - \frac{2 \,c \,\kappa \, \mu}{Z_0^2}\,\right]
\\
\label{constraintPhi} \rule{0pt}{.9cm}
\Phi_y'(y)& = &
\frac{1}{c^2-d^2}
\left[\,\frac{4}{\sin^2\theta_y}\left( A_\psi \cos\theta_y+A_\phi \right)
 - d\,\omega_\phi\,\right]
\end{eqnarray}
where $A_\psi$ and $A_\phi$ are the integration constants.

The equation of motion coming from the variation of $Z$ will give us the following constraint for the constant $Z_0$
\begin{equation} \label{constraintZ0}
A_T \, A_V \, Z_0^4 \, - \, A_V^2 \, Z_0^6 \, + \, c \, \kappa 
\left(\alpha  \, c \, - \, 2 \, A_{\psi} \, \mu \right) \, = \, 0 \, . 
\end{equation}
 
Substituting the expressions \eqref{solutionT},  \eqref{constraintV}, \eqref{constraintPsi} \& \eqref{constraintPhi} into the 
first Virasoro constraint \eqref{Virasoro}, we obtain the following expression for $\theta_y'(y)$ 
\begin{eqnarray} \label{constrainttheta}
(\theta_y')^2& = & \frac{4}{(c^2-d^2)^2}
\left[ \frac{c^2+d^2}{d}\,\big(A_\psi\,\omega_\psi+A_\phi\,\omega_\phi - \kappa \, A_T + \alpha \, A_V \big) - 
A_V Z_0^2 \, \left(A_V Z_0^2 - 2 \, A_T \right)
\right. \nonumber \\
\rule{0pt}{.75cm}
&-&\frac{c^2}{4}\,\Big(\omega_\psi^2-2\,\omega_\psi \,\omega_\phi \,\cos\theta_y
+ \omega_\phi^2  - 4 \, \kappa^2 + \frac{8 \, \alpha \, \kappa}{Z_0^2} \Big) 
- c \, \mu \, A_V \left( \omega_{\psi} -\omega_{\phi} \cos \theta_y \right)
\nonumber\\
\rule{0pt}{.75cm}
&-&
\left. \frac{4}{\sin^2\theta_y}\left(A_\psi^2
+ 2\,A_\psi\,A_\phi \cos\theta_y +A_\phi^2\,\right) - \mu^2 \,A_V^2 + \frac{4 \, c\, \kappa \, \mu}{Z_0^2} \, A_{\psi}  
\,\right].
\end{eqnarray}
Using \eqref{constrainttheta} it is possible to prove that the equation of motion coming from the variation along 
the direction $\theta$ is satisfied. Finally, substituting the expressions \eqref{solutionT},  \eqref{constraintV}, 
\eqref{constraintPsi},  \eqref{constraintPhi} \& \eqref{constrainttheta} into the second Virasoro constraint \eqref{Virasoro}, 
we obtain the following algebraic relation
\begin{equation}
\label{virasoro}
A_\psi\,\omega_\psi+A_\phi\,\omega_\phi - \kappa \, A_T + \alpha \, A_V =0 \, .
\end{equation}
From the analysis above it is clear that once we solve \eqref{constrainttheta} we can substitute the solution for $\theta_y(y)$
in equations \eqref{constraintV}, \eqref{constraintPsi} and \eqref{constraintPhi}, in order to integrate them and obtain $V_y(y)$,
$\Psi_y(y)$ and  $\Phi_y(y)$ respectively. 

The next step in the direction of solving \eqref{constrainttheta} is to introduce a new function, namely $u(y)$, as follows
\begin{equation} \label{def-u}
u \equiv \cos^2 \left(\frac{\theta_y}{2}\right) \, .
\end{equation}
Using \eqref{constrainttheta} we can write the differential equation for $u(y)$
\begin{equation} \label{equation-u} 
\frac{(u')^2}{2}\,+\,{\cal W}(u) \,=\,0
\hspace{.7cm} \textrm{with} \hspace{.7cm}
{\cal W}(u)\,=\,-\,2 \left(\beta_6\, u^3 + \beta_4\, u^2+ \beta_2\, u\,+ \beta_0 \right)
\end{equation}
where
\begin{eqnarray}
\label{beta 6}
\hspace{-.7cm}
\beta_6 \,& = & \, - \,  \frac{c \, \omega_{\phi} \big( c \, \omega_\psi + 2 \, A_V \, \mu \big)}{(c^2-d^2)^2}
\\
\label{beta 4}
\hspace{-.7cm}
\beta_4 \,& = & \, \frac{1}{4 \,(c^2-d^2)^2}  \left[
- \frac{4}{d}\,\left(c^2+d^2\right)\,\big(A_\psi\,\omega_\psi+A_\phi\,\omega_\phi - \kappa \, A_T + 
\alpha \, A_V \big)  \right.
\\
\hspace{-.7cm}
&+&  \, c^2\left( \, \omega_\psi^2 +6 \,\omega_\phi\,\omega_\psi+ \omega_\phi^2 - 
4 \, \kappa^2 + \frac{8 \, \alpha \, \kappa}{Z_0^2}\,\right) + 4\,  A_V\, c\, \mu \left(\omega_{\psi}+3 \, \omega_{\phi}\right)
\nonumber \\
\hspace{-.7cm}
&+& \left. \, 4 \, \mu^2 \, A_V^2 + 4 \, A_V \, Z_0^2 \, \left(A_V Z_0^2 - 2 \, A_T \right) - 
\frac{16\, c\, \kappa \, \mu \, A_{\psi}}{Z_0^2} \right]
\nonumber \\
\label{beta 2}
\hspace{-.7cm}
\beta_2 \,& = & \, \frac{1}{4 \,(c^2-d^2)^2}  \left[
\frac{4}{d}\,\left(c^2+d^2\right)\,\big(A_\psi\,\omega_\psi+A_\phi\,\omega_\phi - \kappa \, A_T + 
\alpha \, A_V \big)  \right.
\\
\hspace{-.7cm}
&-&  \, c^2\left( \, \left(\omega_\psi + \omega_\phi \right)^2 - 
4 \, \kappa^2 + \frac{8 \, \alpha \, \kappa}{Z_0^2}\,\right) -  4\,  A_V\, c\, \mu \left(\omega_{\psi}+ \omega_{\phi}\right)
\nonumber \\
\hspace{-.7cm}
&-& \left. \, 4 \, \mu^2 \, A_V^2 - 4 \, A_V \, Z_0^2 \, \left(A_V Z_0^2 - 2 \, A_T \right) + 
\frac{16\, c\, \kappa \, \mu \, A_{\psi}}{Z_0^2}  - 16\, A_{\phi} \, A_{\psi} \right]
\nonumber \\
\label{beta 0}
\hspace{-.7cm}
\beta_0 \,& = & \,   -\,\frac{(A_\psi - A_\phi)^2}{(c^2-d^2)^2} \, .
\end{eqnarray}
Notice that $u(y) \in [0,1]$.


\subsection{Boundary conditions}
\label{BC}

The next important step in order to solve equation \eqref{equation-u} is to impose the correct boundary conditions. 
Here we are interested in a string configuration that will extend between $\theta = \pi$ which defines an equator of $S^3$ and some other angle $\theta_0$ 
that will be determined dynamically from the equation of motion for $\theta$ (or equivalently $u$).  From the 
requirement of finiteness  for $\Psi_y'$, $\Phi_y'$ at $\theta=\pi$ we are led to the following first constraint 
\begin{equation}
\label{constraint-1}
A_\psi \,=\,A_\phi \, .
\end{equation}

Requiring the function $T_y$ from equation \eqref{solutionT} to vanish, we fix the constant $A_V$
\begin{equation}
\label{constant-AV}
A_V\,=\,\frac{d\, \kappa}{Z_0^2}\, .
\end{equation}  

Since the $\theta=\pi$ is an end point for the string configuration, 
we impose that $\theta_y'$ and $V_y'$ at $\theta= \pi$
are zero and not just finite (as $\Psi_y'$, $\Phi_y'$). Combining those two conditions with \eqref{constraintZ0} and
\eqref{constraint-1}  it is possible to fully determine the three remaining constants, namely $A_T$, $Z_0$ and $A_{\phi}$. 
The expressions for the constants $A_T$ and $Z_0$ are the following
\begin{eqnarray} \label{ATandZ0}
&& A_T \, = \, \frac{1}{2 \, Z_0^4} \Bigg[\, c \, \mu  \, Z_0^2 \left(\omega_\psi \, + \, \omega_\phi \right) \, + \, 
2 \, d \left(\kappa  \, \mu^2 \, + \, \kappa \, Z_0^4 \, -\alpha \, Z_0^2 \right) \Bigg]
\nonumber \\[4pt] 
&& Z_0^2 \, = \, \frac{2 \, d^2 \, \kappa  \, \mu^2}{c\, \mu \, \Big[4 \, A_\phi \, - \, d \, 
\left(\omega_\psi \,  + \, \omega_\phi\right) \Big] \, - \, 2 \, \alpha \, c^2 \, + \, 2 \, \alpha \, d^2} \, . 
\end{eqnarray}
Finally substituting in the condition $\theta_y'(\pi)=0$ the value of the constants $A_V,\,A_T$ and $Z_0$ we end up with a second order algebraic equation for $A_\phi$ with solutions that 
characterize the classical string configurations of the giant magnon and the single spike 
\begin{eqnarray}
\label{AGM}
& &
A_\phi\,=\,\frac{1}{4 \, \mu}\,\Big[ 2 \, c \, \alpha \, + d\, \mu \, \left(\omega_\psi \, + \, \omega_\phi \right) \Big]
\hspace{1.8cm} \textrm{giant magnons}
\\
\label{ASS}
\rule{0pt}{.9cm}
& &
A_\phi\,=\,\frac{c\, \alpha}{ 2 \, \mu}
\hspace{6cm} \textrm{single spiky strings} \, . 
\end{eqnarray}
Combining the values of the constants we have already determined and the Virasoro constraint 
\eqref{virasoro}, we calculate the 
values of $\kappa^2$ for the giant magnon and the single spike
\begin{eqnarray}
\label{k2GM}
& & 
\kappa^2\,=\,\frac{4 \, \alpha ^2 \, + \, \mu ^2 \left(\omega_\psi\,  + \, \omega_\phi \right)^2}{4 \,\mu^2}
\hspace{2cm}
\textrm{giant magnons}
\\
\label{k2SS}
\rule{0pt}{.9cm}
& &
\kappa^2\,= \,\frac{\alpha^2}{\mu^2}
\hspace{5.25cm}
\textrm{single spike strings}\,.
\end{eqnarray}
It should be mentioned that taking the limit $Z_0 \rightarrow \infty$ from \eqref{ATandZ0} (that corresponds to the 
limit of zero deformation $\mu \rightarrow 0$) and using the
corresponding expression for the $A_{\phi}$ (\eqref{AGM} for the magnon and \eqref{ASS} for the spike) we 
determine the value of the  constant $\alpha$. Substituting this value in \eqref{AGM} and \eqref{ASS} we obtain the 
value of the constant $A_{\phi}$ in the undeformed case that is analyzed in \cite{Benvenuti:2008bd}
\begin{eqnarray}
\label{undeformedGM}
& &
\alpha \, = \,  0
\quad \Rightarrow \quad A_\phi\,=\,\frac{1}{4}\,d\,(\omega_\psi+\omega_\phi )
\hspace{3.5cm}
\textrm{giant magnons}
\\
\label{undeformedSS}
\rule{0pt}{.9cm}
& &
\alpha \, = \,  \frac{c \,\mu \, \left(\omega_\psi \, + \, \omega_\phi \right)}{2 \, d} 
\quad \Rightarrow \quad 
A_\phi\,=\,\frac{1}{4}\;\frac{c^2\,(\omega_\psi+\omega_\phi )}{d}
\hspace{.95cm}
\textrm{single spike strings} \, .
\end{eqnarray}

Substituting \eqref{constraint-1}, \eqref{constant-AV}, \eqref{ATandZ0}, \eqref{AGM} and \eqref{ASS}
in the definitions for the coefficients $\beta_0$ \eqref{beta 0} and $\beta_2$ \eqref{beta 2} of the potential ${\cal W}$ 
in  \eqref{equation-u} we conclude that 
\begin{equation} \label{beta0-beta2}
\beta_0 \, = \, \beta_2 \, = \, 0 
\end{equation}
while the coefficients $\beta_4$ \eqref{beta 4} and  $\beta_6$ \eqref{beta 6} obey the following relation
\begin{equation} \label{beta4-beta6}
\beta_4 \, + \, \beta_6 \,=\,
- \, \frac{4\,A_\phi^2}{\left(c^2 \, - \, d^2 \right)^2} \, . 
\end{equation}

In the following, since the quantity $c^2-d^2$ is positive for the magnons and negative for the spikes, we 
introduce the quantity $v$ with value $v \equiv d/c$ for the magnons and $v \equiv c/d$ for the spikes. 
Also we define $\Omega \equiv \omega_\psi /\omega_\phi$. 

In order to determine the dispersion relations for the giant magnons and the single spikes we need to solve the 
differential equation \eqref{equation-u}, which after using the boundary conditions reduces to 
\begin{equation} \label{equation-u-new}
\frac{(u')^2}{2}\,+\,{\cal W}(u)
\,=\,0
\hspace{.7cm}
\textrm{with}
\hspace{.7cm}
{\cal W}(u)\,=\,-\,2\,u^2 \left( \beta_4 \,+ \, \beta_6\, u \right) \, . 
\end{equation}


\subsection{Solution and some useful integrals}
\label{solution-integrals}

In this subsection we will discuss the solution of equation \eqref{equation-u-new} with $\beta_4>0$ and $\beta_6<0$, 
in such a way that $\beta_4+ \beta_6 < 0 $ and \eqref{beta4-beta6} is satisfied\footnote{We do not consider the 
cases with $\beta_4 \leq0$ since they do not satisfy the boundary conditions. See also the discussion and analysis 
in \cite{Benvenuti:2008bd} for the zero deformation case.}.  Since the internal sphere $S^3$ is not 
deformed by the presence of $\mu$ the solution is simple and given by the following 
\begin{equation}
\label{u-solution}
u(y)\,=\, \frac{\beta_4}{|\beta_6|}\, \frac{1}{\cosh^2\big(\sqrt{\beta_4}\, y\big)} \, . 
\end{equation}
In the calculation of the conserved quantities for the giant magnon and the single spike we will need the following 
two definite integrals  
\begin{eqnarray}
\label{I1}
{\cal I}_1 & &\equiv\,\int_{-\infty}^{+\infty}u(y)\,dy \,=\, \frac{2 \, \sqrt{\beta_4}}{|\beta_6|}
 \\
\label{I2}
\rule{0pt}{.9cm}
{\cal I}_2 & &\equiv\, \int_{-\infty}^{+\infty}\frac{u(y)}{1-u(y)}\, dy
\, = \, \frac{2}{\sqrt{|\beta_4 \, + \, \beta_6|}} \,\arcsin \left(\sqrt{\frac{\beta_4}{|\beta_6|}} \right) \, .
\end{eqnarray}


\subsection{Angular amplitudes of the solution}
\label{angular-amplitudes}

In order to compute the dispersion relations for the giant magnon and the single spike it is essential to construct and 
compute the angular amplitudes along the directions of $\psi$ and $\phi$. For this we need to integrate equations 
\eqref{constraintPsi} and \eqref{constraintPhi} over $y \in (-\infty, \infty)$. Rewriting those equations in terms of the 
variable $u$ through \eqref{def-u} and using \eqref{constraint-1} we arrive to the following simple expressions
\begin{eqnarray} \label{constraintPsi-new}
& &
\Psi_y' \,=\,
\frac{1}{c^2-d^2}
\left(\,\frac{2 \,A_\phi}{1 \, - \, u} \, - \, d\,\omega_\psi \, - \, \frac{2 \,c \, \kappa \, \mu}{Z_0^2} \right) 
\\
\rule{0pt}{.8cm}
\label{constraintPhi-new}
& &
\Phi_y' \,=\,
\frac{1}{c^2-d^2} 
\left(\,\frac{2\,A_\phi}{1-u} \, - \, d\,\omega_\phi \right) \, .
\end{eqnarray}
From the analysis of the previous subsection \ref{solution-integrals} we know that for the solution $u=u(y)$, given in 
\eqref{u-solution}, the integral of $1/(1-u)$ diverges and therefore both $\Delta\psi$ and $\Delta\phi$ are divergent quantities.
Motivated by the archetypical giant magnon dispersion relation of \cite{Hofman:2006xt}, we  introduce the 
following combination
\begin{equation} \label{Delta-Phi}
\Delta \varphi_1 \,\equiv\,\frac{\Delta \psi \, + \, \Delta \phi}{2}\,=\,
\int_{-\infty}^{+\infty}\,
\frac{\Psi_y' \, + \,  \Phi_y'}{2}\,dy
\end{equation}
that remains finite for the case of the giant magnons while it diverges for the case of the single spikes. 


\section{Giant magnons}
\label{GM}

In this section we calculate the conserved charges for the giant magnon solution and construct the finite 
linear combinations of those conserved quantities that will lead us to the dispersion relation. 
The four conserved charges, coming from the derivatives of the Polyakov action with respect to $\partial_{\tau} T$,
$\partial_{\tau} V$, $\partial_{\tau} \psi$ and $\partial_{\tau} \phi$, for the giant magnon solution 
(that is characterized by the value \eqref{AGM} for $A_{\phi}$) become
\begin{eqnarray}
\label{EnergyGM}
\frac{p_T}{T}&=& \frac{1}{2  \, \mu \, \Omega} \, 
\sqrt{4 \, \alpha^2 \, \Omega^2 \, + \, \mu^2 \, \left(1\, +\, \Omega \right)^2 \, \omega_\psi^2}
\\
\label{MGM}
\rule{0pt}{.8cm}
\frac{p_V}{T} &=&\frac{\alpha}{ \mu^2}
\\
\label{JpsiGM}
\rule{0pt}{.8cm}
\frac{p_{\psi}}{T}
&=& \frac{\left( 1\, + \, \Omega\right) \,\omega_\psi }{4  \, \Omega} \, - \,  
\frac{\omega_\psi}{2  \, \left(1 \,- \, v^2 \right) \, \Omega} \, u(y)
\\
\label{JphiGM}
\rule{0pt}{.8cm}
\frac{p_{\phi}}{T}
&=& \frac{\left( 1\, + \, \Omega\right) \,\omega_\psi }{4  \, \Omega} \,- \,  
\frac{\mu  \, \omega_\psi \, + \, 2 \, \alpha \, v}{2  \, \mu \, \left(1\, - \, v^2\right)} \, u(y)
\\
\label{DeltaGM}
\rule{0pt}{.8cm}
\Delta \varphi_1
&=&  \frac{2 \, \alpha \, \Omega \, + \, \mu \, v \, \left( 1\, + \, \Omega \right) \, \omega_\psi}{2 \,c \, \mu 
\left(1 \, - \, v^2 \right) \, \Omega } \, \,  {\cal I}_2 \, . 
\end{eqnarray}
The four charges\footnote{Here and in the following we use the notation $\mathcal{E} \equiv E/T$, 
$\mathcal{M} \equiv M/T$, $\mathcal{J}_\psi \equiv J_\psi/T$ and $\mathcal{J}_\phi \equiv J_\phi/T$.} 
$\mathcal{E}$,  $\mathcal{M}$, $\mathcal{J}_\psi$ and $\mathcal{J}_\phi$ that are obtained by integrating the momenta in
\eqref{EnergyGM}, \eqref{MGM}, \eqref{JpsiGM} and \eqref{JphiGM}, respectively
diverge because 
of the constant term and only $\Delta \varphi_1$ converges, since the integral in \eqref{I2} converges. 
Taking inspiration from the dispersion relation of the undeformed giant magnon, it is possible to construct 
linear combination of conserved quantities that will lead us to the following dispersion relation
\begin{equation}
\label{dispersionGM}
\left(\sqrt{\mathcal{E}^2 \, - \, \mu^2 \, \mathcal{M}^2} \, - \, \mathcal{J}_1\right)^2-
\mathcal{J}_2^2
\,=\,
4 \sin^2\frac{\Delta \varphi_1}{2}
\end{equation}
where we have used that 
\begin{equation}
\mathcal{J}_1 \, = \, \left(\mathcal{J}_\psi \, + \, \mathcal{J}_\phi \right) 
\quad \& \quad 
\mathcal{J}_2 \, = \, \left(\mathcal{J}_\psi \, - \, \mathcal{J}_\phi \right) \, . 
\end{equation}


\section{Single spikes}
\label{SS}

In this section we calculate the conserved charges for the single spike solution, in the same fashion as we did 
for the giant magnons. The four conserved charges for the single spike solution 
(that is characterized by the value \eqref{ASS} for $A_{\phi}$) become
\begin{eqnarray}
\label{EnergySS}
{p_T \over T} &=& \frac{\alpha }{ \mu \, v}
\\
\label{MSS}
\rule{0pt}{.8cm}
{p_V \over T} &=& \frac{2 \,\alpha \,- \, \mu \, v \, \left(1\, + \, \Omega \right) \, \omega_\phi}{2  \,\mu^2 \, v}
\\
\label{JpsiSS}
\rule{0pt}{.8cm}
{p_{\psi} \over T}
&=&\frac{v \,\omega_\phi}{2  \, \left(1 \, - \, v^2 \right)} \, u(y)
\\
\label{JphiSS}
\rule{0pt}{.8cm}
{p_{\phi} \over T}
&=& \frac{2 \, \alpha \, - \, \mu \, v \, \omega_\phi}{2 \, \mu \, \left(1\,- \, v^2 \right)} \, u(y)
\\
\label{DeltaSS}
\rule{0pt}{.8cm}
\Delta \varphi_1
&=&  \frac{\left( 1\, + \, \Omega \right) \,\omega_\phi }{2 \, d} \, {\cal I}_3 \, 
- \, \frac{1}{2 \, d}\, \left(\frac{2 \, \alpha \, v}{\mu \, \left(1 \, - \, v^2 \right)} \, + \, \left(1\, + \, \Omega \right) \,  
\omega_\phi\right) \, {\cal I}_2 
\end{eqnarray}
where ${\cal I}_3$ in a non-convergent integral\footnote{
That integral is defined as
\begin{equation}
{\cal I}_3 \equiv \int_{-\infty}^{+\infty}\frac{1}{1-u(y)}\, dy
\end{equation} 
and we need to eliminate it by a suitable linear combination of the other conserved quantities.}.
Taking inspiration from the dispersion relation of the undeformed single spike in $S^3$ of \cite{Ishizeki:2007we}, 
it is possible to construct linear combinations of conserved quantities that will lead us to the following dispersion relation
\begin{equation}
\label{dispersionSS}
\frac{1}{4} \, \left(\mathcal{J}^2_1 \, - \, \mathcal{J}^2_2 \right)
\,=\,
\sin^2\left[\frac{1}{2} \, \left(\mathcal{E} \, - \, \mu \, \mathcal{M} - \Delta \varphi_1 \right)\right] \, .
\end{equation}


\section{Discussion}
\label{discussion}

In this paper we have derived semi-classical string solutions living in the Schr\"odinger $Sch_5 \times S^5$ spacetime which is conjectured to be the gravity dual of the non-local CFT coming under the name null dipole CFT. 
Our solutions are the counterparts of the giant magnon and single spike solutions of the undeformed $AdS_5 \times S^5$ background. This claim is supported by the fact that in the limit where the deformation parameter $\mu \rightarrow 0$ our solutions become those of the giant magnon and single spike solutions of the original $AdS_5 \times S^5$ background. The same is true when the solution, and as a result the dual operator, 
does not carry momentum along the $X^{-}$ direction, i.e. $M=0$.  Our solutions live in an $S^3$ subspace of the five-sphere along the directions of which the $B$-field has non-zero components. We have explicitly checked that the submanifold we have used is a consistent sector of classical string theory on $Sch_5 \times S^5$.

The string solutions we have found are open string solutions which have infinite energy and angular momentum. Their end points are situated on the equator of the three-sphere $S^3$ and for the case of the magnon their angle difference $\Delta \varphi_1$ is to be identified with the momentum  $p$ carried by the excitation. In the case of the dyonic giant magnon the dispersion relation is given by 
\begin{equation}
\label{dispersionGM-final}
\sqrt{E^2 \, - \, \frac{\lambda}{4 \, \pi^2} \, L^2 \, M^2} \, - \, J_1\,=\,
\sqrt{J_2^2 \, + \, \frac{\lambda}{\pi^2} \, \sin^2\frac{p}{2}}
\end{equation}
while for the single spike is given by
\begin{equation}
\label{dispersionSS-final}
 \, J^2_1 \, - \, J^2_2
\,=\, \frac{\lambda}{\pi^2} \, 
\sin^2\left[\frac{ \pi}{\sqrt{\lambda}} \, E \, - \, \frac{1}{2}\, L \, M  \, - \, \frac{1}{2} \, \Delta \varphi_1\right] 
\end{equation}
where we have used that 
\begin{equation} \label{mu-def}
\mu \, = \, \frac{\sqrt{\lambda }}{2\pi }\,L
\end{equation}
where $L$ is the parameter defining the $\star$-product in field theory \cite{Guica:2017mtd}. 

Based on \eqref{dispersionGM-final} one may boldly conjecture that the {\it exact in $\lambda$} (in the strict infinite $J\rightarrow \infty$ limit, of course) dispersion relation of a single particle excitation -see also \eqref{magnon1}- in the infinite $J\rightarrow \infty$ limit will be given by \footnote{A more conservative and general but less predictive proposal would be to multiply the $ \sin^2\frac{p}{2}$ by a function depending on the coupling $\lambda$, as well as on the deformation parameter $L$, namely \\
 $\sqrt{E^2 \, - \, \frac{\lambda}{4 \, \pi^2} \, L^2 \, M^2} \, - \, J\,=\,
\sqrt{1 \, + \, \frac{f(\lambda, L)\,\lambda}{\pi^2} \, \sin^2\frac{p}{2}}$. This function should have the asymtotics $f(\lambda=\infty, L)=1$ and should be computed order by order in string or field theory perturbation theory.}
\begin{equation}
\label{dispersionGM-final-2}
\sqrt{E^2 \, - \, \frac{\lambda}{4 \, \pi^2} \, L^2 \, M^2} \, - \, J\,=\,
\sqrt{1 \, + \, \frac{\lambda}{\pi^2} \, \sin^2\frac{p}{2}}, \qquad J\rightarrow \infty.
\end{equation}
In order to be able to compare with the known field theory results, which are available only at one-loop order, one should take the momentum of the excitation small, i.e.  $p=\frac{2 \pi n}{J}$ and the coupling to scale in such a way that the quantity $\lambda' $ will also be small, i.e. $\lambda'=\frac{\lambda}{J^2}<<1$ and expand \eqref{dispersionGM-final-2} around the BMN-like reference state of \cite{Guica:2017mtd}. Keeping only the linear terms in the $\lambda'$ expansion we get
\begin{equation}
\label{dispersionGM-expan}
E \, = \, \left(1 \, + \, J \, + \, \frac{\lambda n^2}{2 J^2}+...\right) \cdot 
\left(1 \, + \, \frac{\lambda\, L^2 M^2}{8\pi^2J^2} \, + ...\right),
\end{equation}
where the dots denote terms of order higher or equal to $\lambda'^2$.
This prediction should be compared to the one-loop result obtained from the Baxter equation which reads \cite{Ouyang:2017yko}
\begin{equation}
\label{dispersionGM-Baxter}
\Delta=E-(J+1)=\frac{\lambda L^2 M^2}{8 \pi^2 J}+\Big( \frac{\lambda n^2}{2 J^2}+\frac{\lambda L M n}{\pi J^2}+\frac{\lambda\, L^2 M^2}{8\pi^2J^2}+\frac{\lambda}{J^2}\,O(L^4M^4)\Big)\,.
\end{equation}
Comparing \eqref{dispersionGM-expan} and \eqref{dispersionGM-Baxter} one easily sees that the string theory prediction correctly reproduces the first, second and fourth term of the field theory result failing to reproduce all other terms. Although \eqref{dispersionGM-final-2} is conjectured to be exact in the $\lambda$ expansion it is valid only in the $J\rightarrow \infty$ limit. We now argue that all the terms in \eqref{dispersionGM-Baxter} that can not be reproduced by the string prediction are actually suppressed by powers of 
$\frac{1}{\sqrt{J}}$. Indeed, in order to be feasible to compare the string result with the field theory perturbative expansion  the first term in \eqref{dispersionGM-Baxter} should be small. For this to happen not only $\frac{\lambda}{J^2}<<1$ but also $L^2 M^2 J\sim 1$. This can be seen by multiplying and dividing the first term of \eqref{dispersionGM-Baxter} by $J$ in order to express it terms of the effective coupling $\frac{\lambda}{J^2}$. Now the condition  $L^2 M^2 J\sim 1$ implies that the quantity $LM$ should scale as $LM\sim \frac{1}{\sqrt{J}}$ in the large $J$ expansion. As a result all terms beyond the second term in \eqref{dispersionGM-Baxter} are suppressed by powers of $\frac{1}{\sqrt{J}}$ and as such can never be obtained from \eqref{dispersionGM-final-2} which is inherently valid in the infinite angular momentum limit. It would be interesting to calculate finite size correction to the dispersion relation \eqref{dispersionGM-final-2}.

One is thus temped to interpret the solution with dispersion relation \eqref{dispersionGM-final-2} as the dispersion relation of  single particle excitations which propagate with momentum $p$ along a certain pseudovaccum/reference state, the strong coupling dual of which is the BMN-like point-like string of \cite{Guica:2017mtd}.  The latter resembles the BMN string of the undeformed $AdS_5 \times S^5$ background. In the supegravity limit ($p=0$) our solution becomes point-like and identical to the one presented in \cite{Guica:2017mtd}.

One can further speculate on the form of the field theory operators which are dual to our string solutions. 
As discussed in \cite{Guica:2017mtd}, 
the Schr\"odinger/null-dipole CFT correspondence can have two equivalent descriptions. One can either consider closed string theory with periodic boundary conditions on the Schr\"odinger background or, equivalently, open string theory with twisted boundary conditions on the original $AdS_5 \times S^5$ background. On the field theory side the first choice is to employ conformal operators whose elementary constituents are the same as in $\mathcal{N}=4$ SYM and which transform under local gauge transformations. Their form is as follows
\begin{equation}\label{operators1}
 \mathcal{O}=\mathop{\mathrm{tr}}\hat{\Phi} _1\ldots \hat{\Phi} _J
\end{equation}
where the hatted fields are those obtained after the Seiberg-Witten map and read 
\begin{equation}\label{hatted}
 \Phi (x)={\rm P}\exp\int_{x+L_\Phi}^{x}dv\,A_-(v)\cdot
\, \hat{\Phi }(x)\cdot\,{\rm P}\exp\int_{x}^{x-L_\Phi}dv\,A_-(v).
\end{equation}
The integration of the Wilson lines is along the light ray connecting the points $x$ and $x\pm L_\Phi$ and $L_\Phi=\frac{LR_{\Phi}}{2}$ is half of the dipole length of the field $\Phi$ having R-charge $R_{\Phi}$. For more details one can refer to \cite{Guica:2017mtd}.
The corresponding spin chain is  periodic and as a consequence dual to the closed string theory on the deformed background. 

An alternative basis which is more convenient in describing the ground state on which the giant magnon propagates is that of the light-ray 
operators defined by a number of fields located along a light-ray, namely \cite{Belitsky:2004sc}
\begin{equation}\label{magnon1}
 \mathcal{O}(x_1,...,x_J) =\mathrm{tr}\,   \tilde{Z}_{x_1}(0)\ldots \tilde{Z}_{x_J}(0) 
\end{equation}
where we have defined the tilded fields as 
\begin{equation}\label{hatted}
 \tilde\Phi_x (0)={\rm P} \exp {\int_{0}^{x}dv\,A_-(v)}\cdot
\, \hat{\Phi }(x)\cdot\,{\rm P} \exp {\int_{x}^{0}dv\,A_-(v)}.
\end{equation}
Notice that the field $\hat{\Phi }(x)\rightarrow U^{-1}(x)\hat{\Phi }(x)U(x)$ transforms under the gauge transformations as a local  field situated at the space-time point $x$  while the field $ \tilde\Phi_x (0)\rightarrow U^{-1}(0)\tilde\Phi_x (0)U(0)$ transforms as a local  field situated at the space-time point $0$. 

In this basis of operators the ground state dual to the BMN-like string of \cite{Guica:2017mtd} will have the form
\begin{equation}\label{ground}
 \mathcal{O}_0 =\int \prod_{i=1}^J \,dx_i\,\,\, \psi_0(x_1,\ldots ,x_J)\,\mathcal{O}(x_1,\ldots ,x_J)
\end{equation}
for a certain wavefunction $\psi_0(z_1,\ldots ,z_J)$. This state belongs to an $SL(2)$ closed subsector and its eigenvalues for an arbitrary number $J$ of fields $Z$ has been calculated in  \cite{Guica:2017mtd} by using the Baxter equation in this sector.
In what follows, we will assume that the \eqref{ground} has a well-defined limit when the number of scalar fields becomes arbitrary large, i.e. when $J\rightarrow \infty$. Our conjecture is that the giant magnon solutions of this work will be dual to the 
following field theory operator
\begin{equation}\label{magnon}
 \mathcal{O}_{magnon} =\int \prod_{i} \,dx_i\,\,\, \psi_0(\ldots ,x_{l-1},x_l,x_{l+1}\ldots)\,\mathcal{O}_p(\ldots x_{l-1},x_l,x_{l+1}\ldots)
\end{equation}
where 
\begin{equation}\label{magnon1}
 \mathcal{O}_p(\ldots x_{l-1},x_l,x_{l+1}\ldots)=  \sum_l e^{i p l} \Bigg (  \ldots \tilde{Z}_{x_{l-1}}(0) \tilde{\Phi}_{x_{l}}(0) \tilde{Z}_{x_{l+1}}(0))\ldots\Bigg ).
\end{equation}
As discussed below \eqref{hatted} the fields $ \tilde\Phi_x (0)$ transform under gauge transformations as if they were all located at the same point $0$. As a result \eqref{magnon1} has striking similarity with the magnon operator in $\mathcal{N}=4$ SYM where all fields are 
sitting at the same point too. Let us also mention that $l$ in \eqref{magnon1} labels the position of the impurity $\tilde{\Phi}_{x_{l}}(0)$ in the long string of $\tilde{Z}_{x_{i}}(0)$'s and that the space-time points $x_i$ of \eqref{magnon1} used in the definition of the tilded fields \eqref{hatted} are not necessarily ordered along the light-ray on which they are defined.

In this approach 
the scalar field $\hat{\Phi}$ is one of the scalars of the three-sphere $S^3$ and is sitting in the $l^{th}$ site of the infinitely long string of the $Z$'s. A couple of important comments are in order. Firstly, the operator in \eqref{magnon} involves a single excitation $\Phi$ and thus is dual to the string solution with $ J_2=J_{\psi}-J_{\phi}=0$ (see also \eqref{dispersionGM-final}). The operator dual to the dyonic magnon of \eqref{dispersionGM-final} is a bound state of $J_2$ scalar fields propagating coherently with momentum $p$. Secondly, we should stress that unlike the pseudovacuum 
\eqref{ground} the operator of \eqref{magnon1} does not belong to 
the $SL(2)$ subsector and as a result for a generic impurity one can not use the integrability related Baxter equation to calculate its dimension as has been done 
in \cite{Guica:2017mtd} for operators in the aforementioned closed sector. Furthermore, although not apparent the operator of \eqref{magnon1}  contains infinitely many derivatives acting on the scalar fields which are obtained upon Taylor expanding the Wilson lines that are present in the definition of the hatted fields \eqref{hatted}. Finally, in the limit  where one switches the deformation off $\mu\rightarrow 0$ the operator 
\eqref{magnon} becomes the usual $\mathcal{N}=4$ magnon operator since the wavefunction $\psi_0$ becomes a product of $\delta$-functions at the same space-time point.

A number of important questions remain to be answered. It would be interesting to calculate the dimensions of the operators in \eqref{magnon} from the field theory side using integrability techniques or even using Feynman diagrams. 
Additionally, the precise nature of the pseudovacuum on which the excitations propagate should be clarified. 
Furthermore, it would be very interesting to see how integrability of the theory manifests itself in the scattering of the giant magnon solutions we have found in this work. This may shed some light to the validity of the interpretation of our solutions as  single particle asymptotic states whose world-sheet scattering matrix can be subsequently evaluated and compared with the the Drinfeld-Reshetikhin twisted $S$-matrix of the deformed theory. One might be also be tempted to try to use the algebra of the theory in order to compute the exact in the coupling dispersion relation and scattering matrix along the lines of \cite{Beisert:2005tm}. Finally, it would be nice to evaluate finite size correction to the dispersion relation of giant magnons and spikes.


\section*{Acknowledgments}
Centro de F\'isica do Porto is partially funded by FCT through the project CERN/FIS-NUC/0045/2015.

\appendix 

\section{Polyakov action and a consistent truncation}
\label{Polyakov-truncation}

The Polyakov form of the string action is given by the following standard expression\footnote{The string tension $T$ is related 
to the 't Hooft coupling as $T=\frac{\sqrt{\lambda}}{2\pi}$.}
\begin{equation}  \label{Polyakov}
S_P \, = \, - \, {T \over 2} \int d\tau d\sigma \left( \sqrt{-h} \, h^{\alpha\beta} \, g_{\alpha\beta} \,  
- \,  \epsilon^{\alpha\beta} \, b_{\alpha\beta} \right)
\end{equation}
where 
\begin{equation} \label{g&b-definition}
g_{\alpha\beta} \, = \,  G_{MN} \, \partial_\alpha x^M \, \partial_\beta x^N  
\quad \& \quad   
b_{\alpha\beta} \, = \,  B_{MN} \partial_\alpha x^M \, \partial_\beta x^N
\end{equation} 
are the pullbacks of the metric and the $B$-field on the string worldsheet and the tensor density
$\epsilon^{\alpha\beta}$ is defined according to the convention $\epsilon^{01}=1$. 
With $h_{\alpha\beta}$ we denote the worldsheet metric and we choose the conformal gauge where
$h_{\alpha\beta}=\eta_{\alpha\beta}$.  The construction is also supplemented by the Virasoro constraints, 
which are obtained by differentiating \eqref{Polyakov} with respect to $h_{\alpha \beta}$, 
\begin{equation} \label{Virasoro}
G_{MN} \left( \partial_{\tau} x^M \, \partial_{\tau} x^N \, + \,  
 \partial_{\sigma} x^M \, \partial_{\sigma} x^N \right) \, = \,  0
\quad \& \quad 
 G_{MN}  \,\partial_{\tau} x^M \, \partial_{\sigma} x^N\, = \,  0 \, .
\end{equation}
The momentum $p_M$ that is canonically conjugate to the coordinate $x^M$ is given by 
the following expression
\begin{equation} \label{momenta}
p_M\,=\,\frac{\partial \mathcal{L}}{\partial \dot{x}^M}
\end{equation} 
where $ \dot{x}^M \equiv \partial_\tau x^M$.

The 10d $Sch_5\times S^5$ metric is given by the following expression \cite{Guica:2017mtd,Kameyama:2015ufa}
\begin{eqnarray} \label{globalsch5s5}
ds^2&=&- \left(1+ \frac{\mu^2}{Z^4} \right) dT^2  +  \frac{1}{Z^2} \, 
\left(2 dT dV + dZ^2 - \vec{X}^2 dT^2 + d\vec{X}^2 \right) \, 
+ \, ds_{S^5}^2
\nonumber \\
B_2&=&\frac{\mu}{Z^2}\,dT \wedge \left(d\chi + \omega \right) \, . 
\end{eqnarray}
The metric in the five-sphere is written as an $S^1$-fibration over $\mathbb{C}\text{P}^2$
\begin{equation} \label{S5-CP2}
ds^2_{\rm S^5}=\left(d\chi+\omega\right)^2 \, + \, ds^2_{\rm \mathbb{C}P^2}
\quad {\rm with} \quad
ds^2_{\rm \mathbb{C}P^2}= d\eta^2+\sin^2\eta\, 
\bigl(\Sigma_1^2+\Sigma_2^2+\cos^2\eta\,\Sigma_3^2\bigr) \, .
\end{equation}
where the $\Sigma_i ~(i=1,2,3)$ and $\omega$ are defined as follows
\begin{eqnarray}
\Sigma_1&\equiv& \frac{1}{2}(\cos\psi\, d\theta - \sin\psi\sin\theta\, d\phi)
\qquad
\Sigma_2\equiv \frac{1}{2}(\sin\psi\, d\theta + \cos\psi\sin\theta\, d\phi)
\nonumber \\[5pt]
\Sigma_3&\equiv& \frac{1}{2}(d\psi - \cos\theta\, d\phi)
\qquad {\rm and} \qquad
\omega \equiv \sin^2\eta\, \Sigma_3\,. 
\end{eqnarray}
It can be explicitly checked that the following ansatz 
\begin{eqnarray} \label{ansatz-full}
&& 
T\,=\, \kappa \,\tau \, + \, T_y(y)
\hspace{1cm}
V \,=\,\alpha \, \tau \,  + \, V_y(y) 
\hspace{1cm}
Z \,=\, Z_0
\hspace{1cm}
\vec{X} \,= \, \chi \, = \, 0
\nonumber \\ [3pt]
&& \theta\,=\,\theta_y(y)
\hspace{1cm}
\psi\,=\,\omega_\psi\,\tau\,+\,\Psi_y(y)
\hspace{1cm}
\phi\,=\,\omega_\phi\,\tau\,+\,\Phi_y(y)
\hspace{1cm}
\eta \,= \, \frac{\pi}{2}
\end{eqnarray} 
satisfies the equations of motion coming from \eqref{Polyakov} and the 
Virasoro constraints from \eqref{Virasoro}. While the equations of motion for $\vec{X}$ and $\eta$ are 
trivially satisfied, special care should be taken for the equation of motion 
for $\chi$, since in order to be satisfied we need to 
use equations \eqref{constraintPsi} \& \eqref{constraintPhi}. Equipped by the 
preceding analysis we set $\vec{X} =  \chi  = 0$ and $\eta = \frac{\pi}{2}$ in equations 
\eqref{globalsch5s5} and \eqref{S5-CP2} and obtain the 
consistent truncation ansatz of \eqref{globalsch5s3} and \eqref{Bfield}. Finally, we would like to 
stress that interchanging the ansatz for $\chi$ and $\psi$ as follows
\begin{equation} 
\chi\,=\,\frac{1}{2} \left(\omega_\psi\,\tau\,+\,\Psi_y(y) \right)
\quad {\rm and} \quad  
\psi \, = \, 0
\end{equation}
lead us to the same differential equations and dispersion relations.


\end{document}